# Genetic regulation of fluxes: iron homeostasis of *Escherichia coli*

Szabolcs Semsey, Anna M. C. Andersson, Sandeep Krishna, Mogens Høgh Jensen, Eric Massé[1,*] and Kim Sneppen*

Center for Models of Life, Niels Bohr Institute, Copenhagen, Denmark and [1]Département de Biochimie, Université de Sherbrooke, Sherbrooke, Québec, Canada

## ABSTRACT

Iron is an essential trace-element for most organisms. However, because high concentration of free intracellular iron is cytotoxic, cells have developed complex regulatory networks that keep free intracellular iron concentration at optimal range, allowing the incorporation of the metal into iron-using enzymes and minimizing damage to the cell. We built a mathematical model of the network that controls iron uptake and usage in the bacterium *Escherichia coli* to explore the dynamics of iron flow. We simulate the effect of sudden decrease or increase in the extracellular iron level on intracellular iron distribution. Based on the results of simulations we discuss the possible roles of the small RNA RyhB and the Fe–S cluster assembly systems in the optimal redistribution of iron flows. We suggest that Fe–S cluster assembly is crucial to prevent the accumulation of toxic levels of free intracellular iron when the environment suddenly becomes iron rich.

## INTRODUCTION

In the post-genomic era the primary focus of system biology studies is to understand the complex molecular networks coordinating cellular processes. Recent advances in data collection and analysis (1–3) have allowed the construction of genome-scale databases that can be utilized to reconstruct regulatory networks (4,5). However, there are major challenges in building quantitative predictive models of molecular networks of whole cells (6). One of the main challenges is to incorporate the interplay between the many important metabolites and the transcription factors, which in turn governs other proteins that manipulate these metabolites (7).

Here, we use the iron homeostatic system of *Escherichia coli* as a model to study the control of a large flux via a small buffer, a common process in biology, engineering and communications.

Iron is an essential trace element for most organisms. It is a highly versatile prosthetic component present in many key enzymes of major biological processes (8). Iron is the second most abundant metal in the Earth's crust, but it is highly insoluble under aerobic conditions at neutral pH. Furthermore, cells growing under aerobic conditions have to face the toxicity of excessive intracellular iron levels that generate hydroxyl radicals through the Fenton reaction. For these reasons, complex regulatory networks have evolved to keep free intracellular iron within a narrow margin, allowing the incorporation of the metal into iron-using enzymes and minimizing damage to the cell. In addition, iron acquisition is a crucial limiting factor for pathogenic bacteria to colonize the host. Recent progress in understanding regulation of iron homeostasis and quantification of several underlying molecular interactions provides an exciting opportunity for modeling studies.

Bacterial iron homeostasis is best understood in *E.coli*. When intracellular iron level is high, the protein Fur (Ferric uptake regulator) represses transcription initiation of iron uptake genes (9,10). Fur also represses a small RNA (sRNA), named RyhB, which facilitates degradation of the mRNAs encoding for Fe-using proteins. Thus, RyhB is derepressed by Fur under low iron condition and degrades ∼20 mRNAs involved in iron metabolism. This mechanism shares similarities with *Pseudomonas aeruginosa* (11) and *Saccharomyces cerevisiae* (12) [for a review see (13)]

Using a systems approach, we analyze iron uptake and usage in exponentially growing *E.coli* cells under aerobic conditions. We build a mathematical model to study the design architecture and dynamic behavior of the underlying biological network. This study focuses primarily on the response to changes in iron availability. Those elements of the iron homeostatic machinery that are involved in responses to specific conditions, e.g. redox stress, or iron storage during the transition to stationary phase, are not included in the model.

*To whom correspondence should be addressed. Tel: +45 353 25352; Fax: +45 353 25425; Email: sneppen@nbi.dk
*Corresspondence may also be addressed to Eric Massé. Tel: +1 819 346 1110, ext. 15475; Fax +1 819 564 5340; Email: eric.masse@usherbrooke.ca

## MATERIALS AND METHODS

### Mathematical model of the network controlling iron flow

The concentration of loosely bound iron ($Fe_1$) can be expressed as a difference of iron transport and usage (Equation 1).

$$\frac{dFe_1}{dt} = \frac{\beta_{in}}{\tau_g} \times P_{tr} \times \frac{Fe_{out}}{Fe_{out} + K_m}$$
$$- \frac{\beta_N}{\tau_g} \times \frac{Fe_1}{Fe_1 + K_{cut}}$$
$$- \frac{\beta_R}{\tau_g} \times mRNA \times \frac{Fe_1}{Fe_1 + K_{cut}} \quad\quad 1$$
$$- \frac{\beta_I}{\tau_g} \times Isc \times \frac{Fe_1^3}{Fe_1^2 + K_{Isc}^2} - \frac{\beta_S}{\tau_g} \times Suf \times Fe_1.$$

The β values are used to set the size of the terms. $\beta_{in}$ (= 3215 μM) is determined by the iron uptake of *fur* mutants (14), while $\beta_N$ (= 200 μM) is set by the iron content of iron-starved cells (8,15). The parameter $\tau_g$ [= 25/ln2 min, (16)] represents dilution by cell division, which occurs on the timescale of one cell generation. The loosely bound iron pool contains free iron and iron associated with Fur. The first term represents iron influx. It is proportional to $P_{tr}$, which is a collective representation of the iron transport machinery. Iron transport in our model follows the kinetics described for the transport of iron–enterobactin complexes (17) ($K_m$ = 0.394 μM, $V_{max}$ = 6 × 10$^4$/min/cell). For smaller amounts of extracellular iron, $Fe_{out}$, the influx is proportional to extracellular iron. However, at high extracellular iron concentration the iron transport machinery becomes saturated and independent of $Fe_{out}$.

The negative terms represent irreversible fluxes into different pools. The second term represents iron incorporation into proteins that are not regulated by RyhB. The model assumes that the production rate of these proteins is independent of iron concentration. In our model, iron incorporation into these proteins depends on iron concentration only at very low levels of loosely bound iron, represented by $K_{cut}$ (= 0.1 μM). The third term models iron incorporation into RyhB-regulated iron-using proteins. Thus, it is proportional to the concentration of RyhB-regulated mRNAs (*mRNA*). The iron-dependence of iron incorporation into these proteins is assumed to be similar to that of the second term. The last two terms represent iron incorporation into Fe–S clusters. Fe–S clusters are assembled by either the *isc* (fourth term) or the *suf* (fifth term) system. The iron dependence of the Isc mediated Fe–S cluster formation has been chosen to be linear for large $Fe_1$ and becomes cubic when $Fe_1$ falls below $K_{Isc}$ (= 2 μM). The reason for using such a formula is that the *isc* system does not work efficiently at low iron concentrations (18). In contrast, Fe–S cluster formation by the *suf* system is proportional to $Fe_1$ regardless of the size of the loosely bound iron pool. We assume that the concentration of cysteine [0.1–0.2 mM, (19)] is not a limiting factor in Fe–S cluster assembly. For a given $\beta_R$ value, $\beta_I$ and $\beta_S$ were obtained using Equation 1 to fit the steady state condition of wild type and *fur* mutant cells.

The dynamics of several variables ($P_{tr}$, *mRNA*, *Suf* and *Isc*) in the equation for $Fe_1$ also need to be modeled. The dynamics of the transport variable are given by the following equation:

$$\frac{dP_{tr}}{dt} = \frac{1/\tau_g}{1 + (FeFur/K_t)} - \frac{P_{tr}}{\tau_g}. \quad\quad 2$$

The first term represents the production of transport machinery, which is repressed by Fe-bound Fur (*FeFur*). $K_t$ (= 0.55 μM) is the Fe–Fur concentration corresponding to the half-maximal production of the transport machinery. The second term represents the reduction of $P_{tr}$ owing to dilution by cell division, which occurs on the timescale of one cell generation. The level of FeFur depends on $Fe_1$ as well as the total amount of the Fur protein inside the cell [*Fur* = 5 μM, (20)]. The binding of iron to Fur is assumed to happen at much shorter timescales than transcription and translation, and to be in equilibrium. Fur has two Fe binding sites, thus FeFur is obtained by solving the following equation:

$$K_{FeFur}^2 = \frac{(Fe_1 - 2FeFur)^2 \times (Fur - FeFur)}{FeFur}, \quad\quad 3$$

where $K_{FeFur}$ is the dissociation constant of the Fe–Fur complex. The reported values for $K_{FeFur}$ range from 1.2 to 55 μM (21–23). $K_{FeFur}$ = 20 μM was used in this model.

The dynamics of the RyhB-regulated mRNAs encoding Fe-proteins are given by the following:

$$\frac{dmRNA}{dt} = \frac{1}{\tau_g} - \frac{mRNA}{\tau_m} - \frac{\gamma}{\tau_g} \times R \times mRNA. \quad\quad 4$$

The first term represents the production of the mRNAs, which is not regulated. In the second term $\tau_m$ (= 5/ln2 min) represents the passive degradation and dilution of the mRNAs. The last term represents RyhB-mediated degradation (*R* represents RyhB concentration). In this term γ (= 150) is a scaling parameter for the formation of the RyhB–mRNA complex. This unknown parameter does not influence steady state behavior but sets timescale for reaching steady state after iron depletion. The RyhB dynamics are given by

$$\frac{dR}{dt} = \frac{\alpha_R/\tau_g}{1 + FeFur/K_F} - \frac{R}{\tau_R} - \frac{\gamma}{\tau_g} \times R \times mRNA. \quad\quad 5$$

The first term is the production of RyhB, which is repressed by Fe–Fur. In this term, $\alpha_R$ is the ratio of the unregulated levels of RyhB and RyhB-regulated mRNAs, a parameter which influences the flux into non-essential proteins when iron is limited. We set $\alpha_R$ = 4, for which the RyhB-regulated iron flow drops ~50% when extracellular iron is decreased from 17 μM to 0.5 × $K_m$. $K_F$ (= 0.02 μM) is the binding constant of Fe–Fur to its operator site (21). In the second term $\tau_R$ (= 25/ln2 min) represents the passive degradation and dilution of RyhB. The third term represents the active degradation of the RyhB–mRNA complexes, where both RNAs are degraded by RNaseE (24). Using this formula we assume that the amount of RyhB degrading with the *isc* mRNA and mRNAs for Fe–S proteins is insignificant. In Equations 2 and 4 and also later in Equation 6 we express production rates in units of a maximum production rate, as the concentrations in themselves are not important here. However the ratio of maximum production rate of RyhB, to maximum production rate of mRNA is important, and parameterized in term of $\alpha_R$ in Equation 5.

Suf levels are determined by the following equation:

$$\frac{dSuf}{dt} = \frac{1}{\tau_g} \times \left(0.3 + \frac{0.7}{1 + FeFur/K_F}\right) - \frac{Suf}{\tau_g} \quad 6$$

This equation incorporates the observation that low level expression of the *suf* operon was observed in log phase cells in the presence of high Fe–Fur, and *suf* promoter activity increased ∼3-fold in *fur* mutants (25). Because 50 nM Fe–Fur gives complete protection of the Fur binding site in the *suf* promoter region *in vitro*, the basal promoter activity observed *in vivo* is likely not to be due to incomplete binding of Fe–Fur.

The repressor of the *isc* operon, IscR, and the proteins of the *isc* system that catalyze Fe–S cluster assembly are expressed from the same promoter. The *isc* promoter is repressed by the Fe–S–IscR complex. The concentration of this complex (*FeSIscR*) is the solution to the following equation:

$$K_{FS} = \frac{(IscR - FeSIscR) \times (FeS - FeSIscR)}{FeSIscR}. \quad 7$$

The small RNA RyhB mediates the active degradation of the isc mRNA. However, the 5′ region of the *isc* transcript encoding IscR is not affected by RyhB.

IscR levels are governed by the following equation:

$$\frac{dIscR}{dt} = \frac{1\,\mu M \times \alpha_{Isc}/\tau_g}{1 + FeSIscR/K_I} - \frac{IscR}{\tau_g} \quad 8$$

The level of the other Isc proteins is given by

$$\frac{dIsc}{dt} = \frac{\alpha_{Isc} \times mRNA_{Isc}}{\tau_g} - \frac{Isc}{\tau_g}, \quad 9$$

where the iscSUA mRNA level ($mRNA_{Isc}$) depends on the level of the Fe–S–IscR complex and the level of RyhB:

$$\frac{dmRNA_{Isc}}{dt} = \frac{1/\tau_g}{1 + FeSIscR/K_I} - \frac{mRNA_{Isc}}{\tau_m} - \frac{\gamma_I}{\tau_g} \times R \times mRNA_{Isc} \quad 10$$

Finally, the Fe–S level (*FeS*) is given by

$$\frac{dFeS}{dt} = \frac{\beta_S}{\tau_g} \times Suf \times Fe_1 + \frac{\beta_I}{\tau_g} \times Isc \times \frac{Fe_1^3}{Fe_1^2 + K_{Isc}^2} - \frac{\beta_{FeS}}{\tau_g} \times \frac{FeS}{FeS + K_{FSP}} - \frac{FeS}{\tau_g}. \quad 11$$

In this equation *FeS* is produced by a term identical to two of the negative terms in Equation 1, and depleted by two terms representing incorporation of Fe–S clusters into proteins and dilution, respectively. Although RyhB is known to regulate some proteins containing Fe–S clusters, the model uses a constant source of Fe–S using proteins, assuming that the proportion of RyhB regulated Fe–S proteins is insignificant.

We stress that present data are consistent with viewing the regulation of the Fe–S channel as a sub-module of the overall

**Table 1.** Parameters of the Fe–S channel that were used in the simulations

| Parameter | Description | Value |
|---|---|---|
| $K_I$ | Binding constant of the Fe–S–IscR complex to its operator site at the *isc* promoter region | 0.089 µM |
| $K_{FS}$ | Dissociation constant of the Fe–S–IscR complex | 75 µM |
| $K_{FSP}$ | Loosely bound iron concentration at which Fe–S incorporation into proteins is half-maximal | 0.29 µM |
| $\beta_{FeS}$ | Scaling parameter for Fe–S incorporation into proteins | 594 µM |
| $\gamma_I$ | Scaling parameter for the formation of the RyhB–*iscSUA* mRNA complex | 6.12 |
| $\alpha_{Isc}$ | Parameter for Isc protein production | 82 |

iron flow regulation. A sub-module which is essential for making Fe–S, and which in addition supply some additional stability to the steady state behavior of the system at fixed external conditions. Overall we find that our model is robust to the unknown scale of self regulation of the Fe–S pool, as well as moderately robust to the relative weights on the two steps in the IscR self regulation. This robustness is found by analyzing model behavior with respect to parameter choices for the Fe–S system. In particular we find that

(i) One can fit all known data by adjusting $K_I$ from 0.089 to 0.89 µM, and simultaneously making Fe–S–IscR binding 5 times weaker, making 10 times more IscR and increasing the threshold $K_{FSP}$ from 0.29 to 2.9 µM. In effect such an adjustment makes the Fe–S pool larger, and further stabilize the steady state of the overall system (e.g. against reduction in RyhB production).
(ii) Also in the sub part of the Fe–S system defined by the negative feedback loop (IscR↔ Fe–S–IscR↔Fe–S–Iscr-Operator→IscR), one may weaken complex formation by, for example, a factor of 5 while strengthening operator binding by a factor of 1.8, while still fitting the activity of the *isc* promoter for wt, *fur*, *iscR* and *iscS* mutants. Also, such a weakening of Fe–S–IscR binding stabilizes the steady state against a substantial reduction in RyhB production ($\alpha_R$ can be lowered from 4 to 0.5).

The parameters of the Fe–S channel that were used in the simulations are summarized in Table 1.

## RESULTS

### The *E.coli* iron network

In this section we describe those elements and interactions of the *E.coli* iron homeostatic system that we used for the construction of the mathematical model. The map of the regulatory network, illustrated with the iron flow, is presented in Figure 1.

When growing exponentially under aerobic conditions, an *E.coli* cell contains ∼1.2 × 10$^6$ atoms of iron (14). However, only ∼1% of these atoms (∼10$^4$) can be found in the free or loosely bound state (26). The size of the available intracellular iron pool is sensed by the ferric uptake regulator (Fur) protein. Fur is a dimeric protein present in ∼5000 copies in log phase cells (20). Fur, when bound by iron (Fe–Fur), represses transcription of several genes for iron uptake (9,10) by binding to a 19 bp sequence (Fur box) in their promoter region. The affinity of Fe–Fur to the 19 bp consensus binding site has been determined (21) ($K_d$ = 20 nM). Fur preferentially binds $Fe^{2+}$, however, the reported binding affinities vary [$K_d$ of $1.2 \times 10^{-6}$ to $5.5 \times 10^{-5}$ M, (21–23)]. Derepression of iron transport in *fur* mutants results in about a 2.4-fold increase in intracellular iron concentration (14), and about a 7-fold increase in the size of the loosely bound iron pool (26). Fe–Fur can also indirectly activate some genes encoding iron-using proteins (27). A close observation of the mechanism of Fur activation revealed that Fe–Fur represses transcription of a small regulatory RNA, RyhB, which promotes degradation of several mRNAs encoding for non-essential iron-using proteins (28,29). RyhB expression increases 40-fold in cells grown in iron depleted LB compared with normal LB (30). When expressed, RyhB binds its target mRNAs and recruits the RNA degradosome, which simultaneously degrades both RyhB and the mRNA (24). Conversely, in iron-replete conditions, active Fur represses RyhB and allows the mRNA targets to be expressed (24). However, this mechanism of indirect Fur activation is not universal in bacteria. Fe–Fur can directly activate transcription by binding close to promoter regions of genes encoding for iron-using (in *Neisseria meningitidis*) or iron storage (in *Helicobacter pylori*) proteins (31,32).

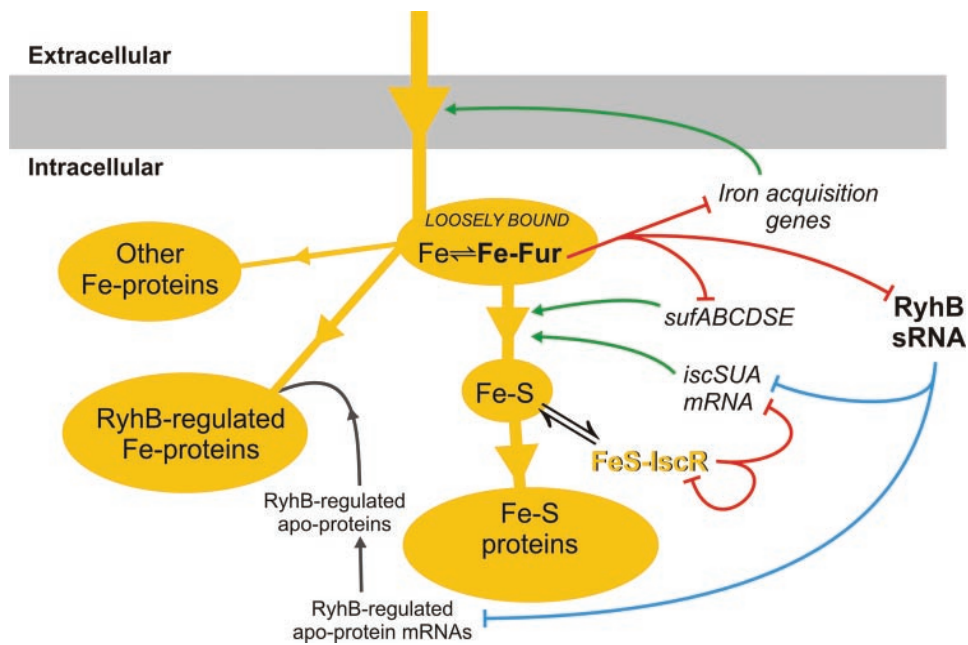

**Figure 1.** Schematic model of iron flow control. Deep yellow arrowheads indicate iron flow; green arrows indicate positive effects on iron flow. Red lines indicate inhibition of transcription, while blue lines indicate inhibition of translation (via mRNA degradation). Iron pools distinguished in the model are shown in deep yellow.

In metalloproteins of *E.coli*, iron can be present as an isolated ion (e.g. in SodB), or can be coordinated with a non-protein organic compound (e.g. in hemoproteins) or a non-metallic ion (e.g. in iron–sulfur proteins). In the model we distinguish iron (Fe) and iron–sulfur (Fe–S) proteins. Proteins belonging to either of these groups can be independent of or regulated by the sRNA RyhB. However, in the model we assume that the fraction of RyhB-regulated Fe–S proteins is negligible.

Assembly of iron–sulfur clusters in *E.coli* is divided into two systems, *isc* and *suf* (33). Although these two systems have overlapping functions, they act as distinct complexes (18). Single operons for both the *isc* and the *suf* systems are derepressed under iron starvation, however, by different mechanisms. The *suf* operon (encoded by *sufABCDSE*) is repressed by Fe–Fur. Expression of the *suf* operon is relatively weak in log phase cells growing in LB, and is increased about 3-fold in *fur* mutants (25). Binding of Fe–Fur to the *suf* promoter region has also been demonstrated *in vitro* (18). In contrast to *suf*, the *isc* operon (encoded by *iscRSUA*) is regulated by a feedback mechanism including Fe–S. Excess amount of Fe–S clusters is sensed by the IscR repressor. Fe–S clusters form a complex with IscR, which then binds to the *isc* promoter to inhibit *isc* transcription (34). The *isc* promoter is derepressed ~17-fold in *iscR* mutants and ~11-fold in *iscS* mutants, which are defective in Fe–S synthesis (34). About 38-fold repression of the *isc* promoter activity was observed *in vitro* in the presence of Fe–S cluster containing IscR. In contrast to a microarray study that reported no effect of *fur* mutation on *iscRSUA* transcript (35), recent results demonstrated that the *isc* transcript is subject to partial RyhB-mediated degradation (28). Indeed, RyhB seems to acts specifically on *iscSUA* transcript, located downstream of *iscR*.

The physiological roles of the *suf* and *isc* systems are divergent. In cells grown in LB during log phase the *isc* system plays the housekeeping role. Despite the induction of the *isc* operon under iron starvation conditions, Fe–S biosynthesis by the *isc* system is not efficient when iron availability is limited (18). However, the *suf* system is adapted for Fe–S biosynthesis under iron starvation conditions. It can efficiently compete for the limited free iron pool, and is probably more successful in protecting sulfane sulfure from loss when iron is unavailable (18). We assume that most of the Fe–S clusters are assembled by the *isc* system under the conditions we simulate.

## Construction of a mathematical model of the iron network

We have built a mathematical model based on a system of differential equations that describes the dynamics of the network controlling iron flow. The model is described in Materials and Methods. The model was designed to simulate the effect of perturbations in the level of available extracellular iron on the iron flow in cells during exponential growth in complete media. Therefore two major components of iron homeostasis, redox stress response and storage, are omitted from the model. Most probably iron storage is negligible under the conditions we simulate, because iron content of logarithmic phase cells is not affected by mutations in *ftnA* and *bfr*, encoding iron storage proteins ferritin and bacterioferritin, respectively (15). In *E.coli*, ferritin is responsible for iron accumulation and storage, <1% of the total cellular iron is bound to bacterioferritin (36). However, *ftnA* is expressed at very low level in logarithmic phase cells grown in LB (29), and induced in the post-exponential growth phase under iron rich conditions (15,37). We use a constant 25 min generation time in the simulations; the model was not designed to simulate changes in cellular growth rate upon changes in iron availability.

## Kinetic responses of the network to rapid changes in the available extracellular iron concentration

In order to achieve steady state intracellular iron concentration during exponential growth, every cell must maintain a regular flow of iron into each pools of the metal (Figure 1). Iron influx approaches its maximum ($1.2 \times 10^6$ Fe atoms/cell generation) at ~1 µM extracellular iron concentration (17). At this influx the loosely bound iron pool in each cell contains ~$10^4$ Fe atoms (~0.8%) (26). In our model there is a constant iron flow to proteins that are not regulated by RyhB ('Other proteins'). We define the size of this pool as the iron content of iron-starved cells [$2 \times 10^5$ atoms (8,15)]. Iron flows to the RyhB-regulated proteins and to Fe–S cluster containing proteins are regulated according to iron availability. Iron flow to the RyhB-regulated proteins is proportional to the expression of these proteins, while Fe–S cluster assembly is proportional to both the concentration of loosely bound iron and the level of the Isc and Suf enzymes.

We simulated the effect of rapid changes in the available extracellular iron concentration on the iron flow in wild-type cells. In Figure 2 we show the effect of decreasing the extracellular iron concentration from 17 (iron concentration in LB) to 0.2 µM (50% of $K_m$ for iron transport) at 0 time point. We performed the simulations at different ratios of the iron flows to the RyhB-regulated proteins and to the Fe–S clusters because there are no experimental measurements available. The decreased iron influx resulted in a similar drop in the loosely bound iron pool, regardless of the size of the RyhB-regulated iron pool (Figure 2). However, the proportion of the RyhB-regulated iron pool greatly affects the allocation of the limited amount of available iron.

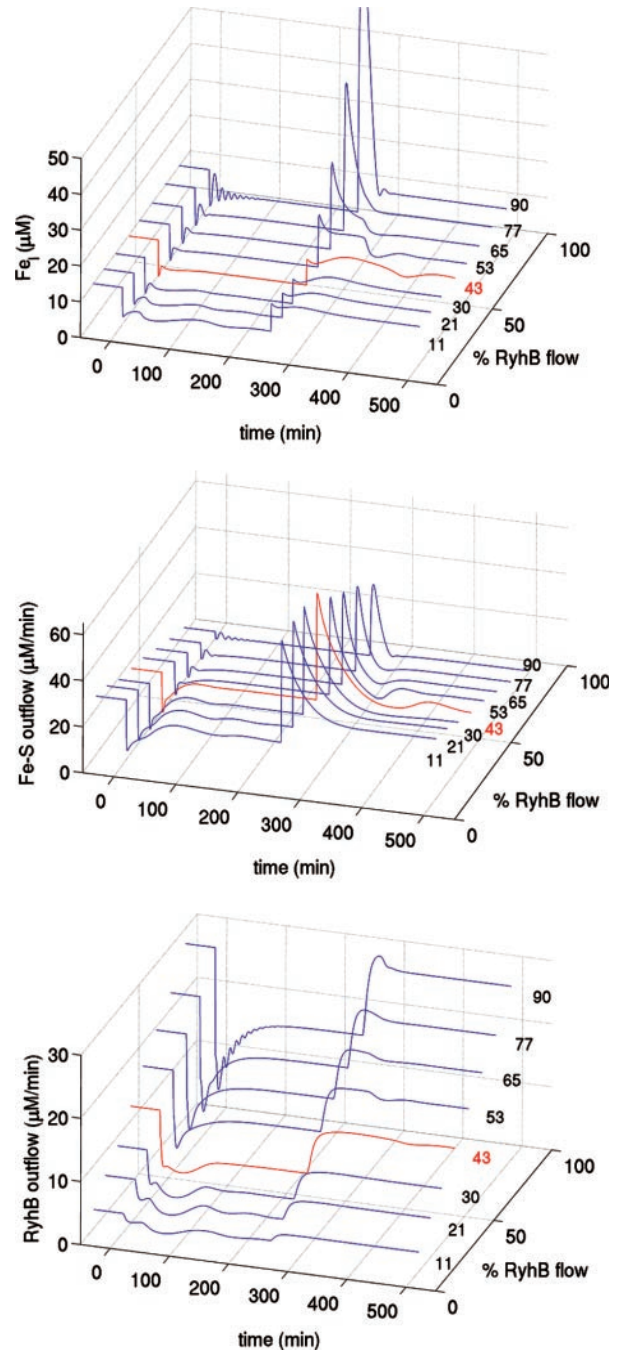

**Figure 2.** Effects of perturbations in the available extracellular iron on the loosely bound iron pool (top), on Fe–S cluster assembly (middle), and on iron flow to the RyhB-regulated Fe-proteins (bottom). Extracellular iron concentration was decreased from 17 to 0.2 µM at 0 time. At the 250 min time point the 17 µM iron concentration was restored. Simulations were performed at different initial distributions of the regulated iron fluxes, indicated by the ratio of the iron flow to the RyhB-regulated Fe-proteins at 17 µM extracellular iron (% RyhB flow). The flux parameters for our standard simulation (red curve, see also Figure 3) are the following: $\beta_{in} = 3215$ µM, $\beta_N = 200$ µM, $\beta_R = 2530$ µM, $\beta_I = 58$ and $\beta_S = 34$.

At lower sizes of the RyhB-regulated iron pool, Fe–S cluster assembly (predominantly by the *isc* system) was maintained close to the original level, while iron flow to the RyhB-regulated Fe-proteins decreased more dramatically (Figures 2 and 3). Redirection of iron to the Fe–S channel upon iron limitation was most efficient when initially ~40% of the regulated iron flow was allocated to RyhB-regulated proteins at 17 μM extracellular iron concentration (Figure 3). This maximum did not depend on the extent of the drop in the extracellular iron concentration (data not shown). To explore how this effect is achieved, we studied the level of *isc* mRNA at different iron concentrations. Transcription of the *iscRSUA* operon is negatively regulated by the Fe–S–IscR complex. However, RyhB mediates active degradation of the 3′ region of the mRNA containing *iscSUA*. The effect of this dual regulation on the *isc* transcript level is shown in Figure 4. The model predicts highest expression of the Isc proteins at intermediate iron availability. When the level of available extracellular iron is further decreased, RyhB reduces the amount of the *iscSUA* transcript. However, the *iscSUA* transcript is only partially degraded by RyhB, despite the fact that the *isc* system is not functional during iron starvation conditions (18). The predicted level of the Isc proteins is higher at lower iron concentrations compared to when iron is not limiting.

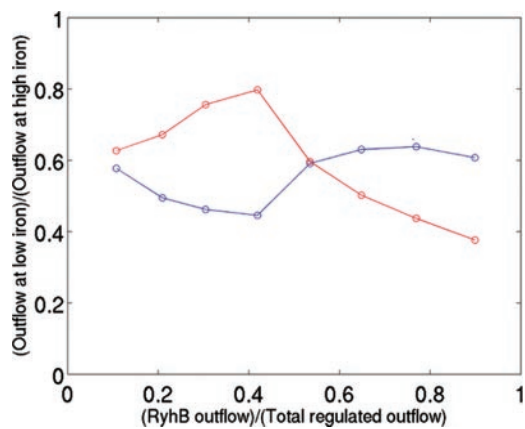

**Figure 3.** Effect of the initial distribution of the regulated iron fluxes on the steady state flows at decreased iron availability. Simulations were performed at different initial distributions of the regulated iron fluxes, indicated by the ratio of the iron flow to the RyhB-regulated Fe-proteins at 17 μM extracellular iron (RyhB outflow/Total regulated outflow). Steady state iron flows to Fe–S cluster assembly (red) and to the RyhB-regulated Fe-proteins (blue) at 0.2 μM extracellular iron were plotted. Values are normalized to the corresponding flows at 17 μM extracellular iron. We used the initial distribution of fluxes corresponding to the maximum of the red curve as a standard condition for simulations (see also Figure 2).

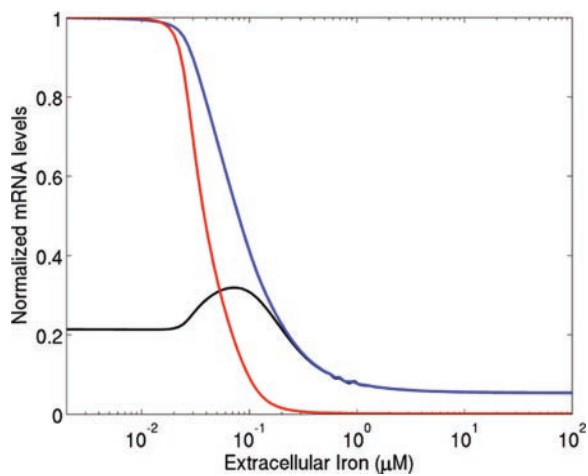

**Figure 4.** The amount of *isc* transcript as a function of iron concentration. RyhB and *iscR* RNA levels at low iron concentration were used as a reference (=1) to normalize RyhB and *isc* RNA levels, respectively. The black curve shows the level of iscSUA transcript which is regulated by both the Fe–S–IscR complex and the small RNA RyhB. The normalized RyhB level is shown in red. The blue curve represents the amount of the IscR protein.

When the 17 μM extracellular iron concentration was rapidly restored at the 250 min. time point (Figure 2), a sudden increase in iron influx (~$2.6 \times 10^6$ Fe atoms/cell generation) was obtained owing to the increased expression of the iron acquisition genes at 0.2 μM extracellular iron concentration. This pulse was rapidly directed into Fe–S clusters by the *isc* system, and thus the pulse in the size of the loosely bound iron pool is damped. This effect mostly depends on the capacity of the *isc* system, higher proportion of the RyhB-regulated iron pool results in higher levels of loosely bound iron.

## DISCUSSION

### Response to iron perturbations

The basic principle of bacterial iron homeostasis is to maintain the intracellular pool of free iron at a level that supplies sufficient iron for iron-using proteins but is not toxic to the cell. To achieve this balance, iron acquisition and consumption is regulated by a complex network. The network contains two transcriptional regulators, Fur and IscR, responding to the concentration of free iron and Fe–S clusters, respectively. The system gives a complex response to restricted iron influx. The lower level of intracellular free iron results in lower Fe–Fur concentration, enhancing the expression of the iron acquisition genes, the *suf* genes and RyhB. RyhB inhibits the production of non-essential iron-using proteins to increase the iron pool available for essential iron-using proteins (13).

We simulated how the size of the RyhB-regulated iron flow could affect intracellular iron distribution when extracellular iron levels are perturbed. Our simulations suggest that the system's response to iron perturbations depends significantly on the size of the RyhB-regulated iron flow (Figures 2 and 3). There is an 'optimal' distribution of intracellular iron flows, at which the system can best fulfill the requirements of iron homeostasis. At this 'optimal' distribution (red curves in Figures 2 and 3), the system can efficiently maintain iron flow to Fe–S cluster assembly when iron influx is limited, by downregulating the synthesis of iron-using proteins. Furthermore, the size of the loosely bound iron pool is efficiently restored without reaching toxic levels when iron suddenly becomes available. Conversely, when a large portion of iron flow is regulated by RyhB, Fe–S cluster assembly becomes more sensitive to iron availability. Simulations also predict accumulation of higher levels of loosely bound iron during the transition from low to high iron conditions.

### Regulation of Fe–S cluster assembly

At the 'optimal' distribution of iron flows discussed above, the level of the Isc proteins can be efficiently adjusted by the combined action of RyhB and IscR, to meet the cell's needs at different availabilities of iron. Reduced iron influx results in lower levels of loosely bound iron, therefore reduced Fe–S cluster assembly. Because the level of Fe–S-IscR is also reduced, the expression of the Isc proteins is increased to restore the optimal rate of Fe–S cluster assembly (Figures 3 and 4). In iron starvation conditions, the *isc* operon is fully derepressed; however, the Fe–S cluster assembly by the *isc* system becomes inefficient. As the level of loosely bound iron decreases, RyhB becomes derepressed, and mediates partial degradation of the 3′ region of the *isc* transcript, encoding IscS, IscU and IscA. The predicted level of Isc proteins is ∼3-fold higher in iron starvation conditions compared with when iron is not limiting. Our dynamic simulations pinpoint the advantage of the elevated level of the Isc proteins in iron-restricted cells. In these cells the expression of the iron acquisition genes is derepressed, allowing iron influx at high capacity when the environment suddenly becomes iron-rich. We suggest that during this transition period the *isc* system assembles the incoming iron into Fe–S clusters, thus preventing the intracellular free iron level from reaching toxic levels. We also suggest that the advantage in IscR not being regulated by RyhB is that the higher level of IscR allows faster shut-down of *isc* transcription at the transition from low to high iron availability. After this transition period, the reduced level of the Isc proteins allows the redirection of iron flow to the newly synthesized RyhB-regulated iron-using proteins.

### Role of small RNA regulation

The advantages of small RNA regulation over transcriptional regulation have been described recently (24,28). One possible advantage of small RNA regulation in iron flow control is the suboperonic discoordination of protein expression by mediating selective degradation of parts of a polycistronic mRNA (38). This control mechanism may allow selective inhibition of synthesis of iron-using proteins produced from operons also encoding proteins that do not require iron for their function.

The example of the *isc* operon shows that transcriptional and small RNA regulation can be combined in an efficient way to adjust protein levels according to the levels of different metabolites. This system requires less complexity at the promoter region compared to an analogous arrangement based on the combination of transcriptional repression and activation (39).

### Concluding remarks

We have analyzed the iron regulation network as a case study of how a large flux can be regulated through a small buffer. In exponentially growing cells the network partitions the incoming iron flux into three parts, an essential part, a Fe–S pool, and a part associated to Fe usage by non-essential proteins. Although the usage governed through the latter two modules appears largely interconnected (Figure 1), our modeling shows that the relative strengths of the interactions indeed allow us to understand iron regulation in terms of a modular structure with two competing sub-systems. In effect, the organization of the network is hierarchical, with free Fe guiding two regulatory modules, and with all feedback going through changes in Fe due to irreversible absorption of Fe in either of the two pathways. We further saw that the Fe–S pathway responds very fast to changes in iron availability, a feature which effectively protects the cell against Fe poisoning under sudden iron bursts. On the other hand, the fact that Fe–S absorption is also shut down under iron depletion puts strain on the non-essential usage. To avoid using Fe in non-essential proteins, the cell needs to stop production of non-essential proteins as fast as possible. This may be the reason for using small RNA-mediated post transcriptional regulation, which indeed short-circuits the genetic regulation upon sudden iron depletion, by removing all mRNA for non-essential iron-using proteins. However, small RNA regulation in iron homeostasis is not universal in bacteria. Our model provides a framework to study the possible advantages of transcriptional and small RNA regulated post-transcriptional control of non-essential iron-using proteins.

## ACKNOWLEDGEMENTS

The authors thank Ian Dodd and Namiko Mitarai for critical reading, discussion and helpful comments on the manuscript. This work was supported by the Danish National Research Foundation. E.M. is a New Investigator Scholar of the Canadian Institutes for Health Research (CIHR) and is supported by an operating grant MOP69005 from the CIHR. Funding to pay the Open Access publication charges for this article was provided by the Danish National Research Foundation.

*Conflict of interest statement*. None declared.